\documentclass[reprint,aps,prb,amsmath,amssymb,superscriptaddress,floatfix,citeautoscript]{revtex4-1}
\usepackage{amsmath}
\usepackage{graphicx}
\usepackage{color}
\usepackage{multirow}
\usepackage[caption = false]{subfig}
\usepackage{multirow}
\usepackage[table,xcdraw]{xcolor}
\usepackage{mhchem}
\usepackage{bbold}

\usepackage{array}
\usepackage{threeparttable}
\usepackage{mathtools}
\usepackage{siunitx}
\usepackage{centernot}
\usepackage{tikz}
\usepackage[normalem]{ulem}
\useunder{\uline}{\ul}{}

\usepackage{gensymb}
\usepackage[colorlinks=true,allcolors=blue]{hyperref}

\usepackage{braket}

\usepackage{txfonts}

\begin{document}
\title{Nuclear gradients from auxiliary-field quantum Monte Carlo and their application in geometry optimization and transition state search}

\author{Jo S. Kurian}
\affiliation{Department of Chemistry, University of Colorado, Boulder, CO 80302, USA}

\author{Ankit Mahajan}
\affiliation{Department of Chemistry, Columbia University, New York, NY 10027, USA}

\author{Sandeep Sharma}
\email{sanshar@gmail.com}
\affiliation{Department of Chemistry, University of Colorado, Boulder, CO 80302, USA}
\affiliation{Division of Chemistry and Chemical Engineering, California Institute of Technology, Pasadena, California 91125, USA}
\affiliation{Marcus Center for Theoretical Chemistry, Pasadena, California 91125, USA}

\begin{abstract}
In this article, we present a method for computing accurate and scalable nuclear forces within the phaseless auxiliary-field quantum Monte Carlo (AFQMC) framework. Our approach leverages automatic differentiation of the energy functional to obtain nuclear gradients at a computational cost comparable to that of energy evaluation. The accuracy of the method is validated against finite difference calculations, showing excellent agreement. We then explore several machine learning (ML) strategies for learning noisy AFQMC data. These ML potentials are subsequently used to perform geometry optimizations and nudged elastic band (NEB) calculations, successfully identifying the transition state of the formamide–formimidic acid tautomerization. The resulting transition state geometry and barrier heights are in close agreement with coupled-cluster reference values. This work paves the way for highly accurate geometry optimization, molecular dynamics, or reaction path calculations.
\end{abstract}
\maketitle

\section{Introduction}
Modeling potential energy surfaces (PESs) accurately is an important challenge in ab initio electronic structure theory. A good description of energy and nuclear forces is crucial to perform geometry optimization, study reaction pathways, obtain vibrational information, and thermodynamic properties using molecular dynamics~\cite{car1985unified,pulay2014analytical,eckert1997ab,pulay1969ab}.
Density functional theory is often the workhorse in quantum chemistry to perform these calculations because of its favorable cost scaling and ease of computing forces using the Hellmann-Feynman theorem (HFT)~\cite{reveles2004geometry}. However, the PES may not be described correctly owing to the errors in DFT approximations.

Quantum Monte Carlo (QMC) methods offer a robust stochastic framework for solving the many-body Schr\"{o}dinger equation and are well-regarded for their ability to generate highly accurate energies, particularly in strongly correlated systems. Despite these strengths, efficient estimation of nuclear forces has remained a challenging problem because of large bias and variance in force calculations. 
Advancements in the field have led to different estimators to handle these issues,~\cite{assaraf2003zero,badinski2010methods,filippi2000correlated,attaccalite2008stable,moroni2014practical,pathak2020light,motta2018communication,jiang2022general} 
primarily focused on variational (VMC) and diffusion (DMC) Monte Carlo methods. Phaseless auxiliary-field quantum Monte Carlo (ph-AFQMC) is another QMC method that achieves high accuracy for ground state energies with favorable scaling ($O(N^4)$ per sample) and efficient parallelization~\cite{hubbardBenchmark2015,hydrogenBenchmark2017,transitionMetalOxides2020,Shee2020afqmc,Friesner2022afqmc_benchmark,Lee2020FullereneFePorphyrin,sukurma2023benchmark, Sharma2022afqmc,eskridge2019local}. Computation of nuclear forces within the ph-AFQMC framework is less mature. One approach incorporates Pulay corrections within the Hellmann–Feynman formalism and employs a backpropagation technique to compute forces,~\cite{motta2018communication} with a recent extension to solids~\cite{chen2023computation}. In this work, we use reverse-mode automatic differentiation (rev-AD) to evaluate nuclear gradients. By decomposing the energy evaluation into a sequence of elementary operations, rev-AD enables efficient computation of energy gradients at a cost comparable to the original function evaluation, while circumventing the need for evaluating complicated analytical expressions. rev-AD has previously been used in QMC methods to obtain gradients~\cite{sorella2010algorithmic,zhang2023automatic,poole2014calculating} in VMC and DMC, and to obtain response properties in ph-AFQMC~\cite{mahajan2023response}. 

Although ph-AFQMC offers favorable scaling, the evaluation of energies and forces remains computationally expensive, limiting its applicability in practical scenarios. Machine-learned (ML) force fields offer a promising solution by significantly reducing the need for repeated quantum chemical calculations, provided a sufficiently large and representative training dataset is available. In recent years, several ML architectures~\cite{unke2021machine,batzner20223,qiao2020orbnet,unke2019physnet,schutt2017quantum,smith2017ani,frank2022so3krates,unke2021spookynet,ko2021fourth,gasteiger2021gemnet,bartok2010gaussian,deringer2021gaussian,christensen2020fchl,behler2007generalized} have appeared in the literature that can be used to learn the potential energy surface obtained from any underlying method. In most cases, the underlying method is DFT, although QMC data has also recently been used in a variety of applications, such as molecular dynamics and nudged elastic band calculations~\cite{huang2022machine, tirelli2022high, slootman2024accurate,huang2023toward}. A recent article pointed out that the stochastic error in QMC can be used as a resource to reduce the computational cost of generating the data~\cite{ceperley2024training}. For example, by the central limit theorem, one can generate four times as many data points using QMC if each data point is obtained with twice the stochastic error. The additional data appears to lead to a superior fit despite the larger stochastic error. 

In addition to implementing nuclear gradients for ph-AFQMC, we explore its application with the goal of answering the following three questions. First, can we leverage the tradeoff between the amount of AFQMC data and the noise in the data to improve the quality of the fit by using a larger amount of noisy data for a fixed computational budget? Second, given that the evaluation of gradients comes at the cost of added memory and computational overhead, is it preferable to generate a larger amount of energy values rather than a smaller number of energies and gradients? Third, for the small molecules studied in this work, is it preferable to use transfer learning to fine-tune a foundation model UMA (Universal Models for Atoms)~\cite{wood2025umafamilyuniversalmodels} developed by Meta FAIR or is it better to use $\Delta$-learning to fit the energy difference between UMA and AFQMC using Kernel Ridge Regression (KRR) as implemented through the sGDML package~\cite{chmiela2017machine,chmiela2018towards}. The latter scales poorly with system size, and it employs a Coulomb-matrix–like descriptor based on inverse interatomic distances, making it inherently system dependent and non-transferable.

The rest of the article is arranged as follows: in Sec.~\ref{sec:theory}, we review the basic theory of ph-AFQMC and present the theory used to obtain nuclear gradients. For a more detailed discussion on ph-AFQMC, we direct the readers to Ref~\onlinecite{motta2018ab}. In Sec.~\ref{sec:results}, we present the results of test calculations where the theory was first tested against finite difference and other high-level theory methods. We then test different ML methods and use ML models trained on ph-AFQMC gradients to perform NEB calculations to identify the transition state of formamide-formimidic acid tautomerization reaction. 

\section{Theory}
\label{sec:theory}

AFQMC is a projector Monte Carlo method where the ground state is obtained as the asymptotic solution of the imaginary time evolution of a trial state,
\begin{equation}
    e^{-\tau \hat{H}}\ket{\Psi_T} \xrightarrow[]{\tau \rightarrow \infty} \ket{\Psi_0}
\end{equation}
provided the initial trial state is not orthogonal to the ground state. Here $\tau$ is the imaginary time, $\ket{\Psi_T}$ is the trial wavefunction, $\ket{\Psi_0}$ is the ground state and $\hat{H}$ is the ab initio Hamiltonian given by,

\begin{equation}
\begin{aligned}
\hat{H} &= \sum_{pq} h_{pq}\hat{a}_p^\dagger \hat{a}_q
        + \frac{1}{2}\sum_{\gamma pq} \left(L_{pq}^\gamma\hat{a}_p^\dagger \hat{a}_q\right)^2
        + H_0 \\
       &= \hat{H}_1 + \hat{H}_2 + \hat{H}_0
\end{aligned}
\end{equation}

Here, $h_{pq}$ are the modified one-electron integrals, $L^\gamma_{pq}$ are the Cholesky decomposed two-electron integrals in an orthonormal orbital basis, and $H_0$ is a constant. Following Trotterization, the short time propagator can be expressed using Hubbard-Stratonovich transformation as,
\begin{equation}
    e^{-\Delta \tau \hat{H}} = \int dx p(x) \hat{B}(x) + \mathcal{O}(\Delta \tau^2)
\end{equation}
where, $x$ are the auxiliary fields, $p(x)$ is the standard Gaussian distribution and $\hat{B}(x)$ is given as,
\begin{equation}
    \hat{B}(x) = e^{-\Delta\tau \hat{H}_1/2}e^{\sqrt{\Delta \tau }\sum_\gamma x_\gamma \hat{v}_\gamma }e^{-\Delta \tau \hat{H}_1/2}
\end{equation}
with $\hat{v}_\gamma = i\sum_{pq}L^\gamma_{pq}\hat{a}_p^\dagger \hat{a}_q$.
The ground state wavefunction can be estimated stochastically by sampling the auxiliary fields and repeatedly applying the short time propagator sufficiently many times.

Energy sampled from the AFQMC procedure at a particular imaginary time can be expressed as
\begin{equation}
    E = \frac{1}{\sum_W^{N_W}w_W}\sum_{W=1}^{N_W} w_W\frac{\bra{\Psi_T}\hat{H}\ket{\Psi_W}}{\braket{\Psi_T|\Psi_W}},
\end{equation}
where $\ket{\Psi_T}$ is the trial wavefunction. $\ket{\Psi_W}$ and ${w_W}$ are the walkers and their corresponding weights. The energy $E$ is therefore a function of the modified one-electron integrals $h_{pq}(R)$, the Cholesky decomposed two-electron integrals $L^\gamma_{pq}(R)$, the trial $\Psi_T$, walker wavefunctions $\Psi_W$, its weight $w_W$, and the constant $H_0(R)$. 
To control the phase problem inherent in QMC methods, we employ the phaseless approximation~\cite{zhang2003quantum}, which stabilizes the stochastic propagation but introduces a systematic bias in the sampled wavefunction. The extent of this bias can be controlled by using a more accurate trial wavefunction.

The nuclear gradient can then be evaluated as,

\begin{equation}\label{eq:gradient}
\frac{dE}{dR}
=
\sum_{pq}\frac{dE}{dh_{pq}}\frac{dh_{pq}}{dR}
+
\sum_{\gamma pq}\frac{dE}{dL_{pq}^{\gamma}}\frac{dL_{pq}^{\gamma}}{dR}
+
\frac{dH_0}{dR}
\end{equation}
where,
\begin{equation}
\begin{aligned}
\frac{dE}{dh_{pq}} &=
\frac{\partial E}{\partial h_{pq}}
+ \frac{\partial E}{\partial \Psi_T}\frac{\partial \Psi_T}{\partial h_{pq}}
+ \sum_w\left(
\frac{\partial E}{\partial \Psi_w}\frac{\partial \Psi_w}{\partial h_{pq}}
+ \frac{\partial E}{\partial w_w}\frac{\partial w_w}{\partial h_{pq}}
\right), \\
\frac{dE}{dL_{pq}^\gamma} &=
\frac{\partial E}{\partial L_{pq}^\gamma}
+ \frac{\partial E}{\partial \Psi_T}\frac{\partial \Psi_T}{\partial L_{pq}^\gamma}
+ \sum_w\left(
\frac{\partial E}{\partial \Psi_w}\frac{\partial \Psi_w}{\partial L_{pq}^\gamma}
+ \frac{\partial E}{\partial w_w}\frac{\partial w_w}{\partial L_{pq}^\gamma}
\right)
\end{aligned}
\end{equation}

Reverse-mode automatic differentiation provides an efficient method for computing these derivatives with the same scaling as that of obtaining the energy itself. rev-AD is performed over the ph-AFQMC procedure, and the partial derivative of the energy with respect to $h_{pq}$ and $L^\gamma_{pq}$ are stored after each block.
Here, the terms $\frac{\partial E}{\partial \Psi_T}\frac{\partial \Psi_T}{\partial h_{pq}}$ and $\frac{\partial E}{\partial \Psi_T}\frac{\partial \Psi_T}{\partial L_{pq}^\gamma}$ contain the response of the trial wavefunction to the Hamiltonian elements.
These are incorporated by employing a Hartree-Fock solver within the ph-AFQMC framework. Similarly, the response of walkers and their weights are accumulated over the ph-AFQMC steps, which are efficiently computed using rev-AD. The derivative of Hamiltonian elements ($\frac{dh_{pq}}{dR}$, $\frac{dL_{pq}^\gamma}{dR}$) with respect to the coordinates are obtained using central difference with $dR = 10^{-5}$. 

Electronic fluctuations remain qualitatively similar in atom-like local orbitals for small changes in atomic positions. Therefore, in order to have a consistent representation of the integrals with the change in geometry, we use L\"{o}wdin orthonormalized atomic orbitals. 
\begin{equation}
\phi_i(r;R) = \sum_\alpha [S(R)]_{\alpha i}^{-\frac{1}{2}} \chi_\alpha (r;R)
\end{equation}
which are obtained from the underlying atomic orbitals $\{\chi_\alpha (r;R)\}$. Here $S(R)$ is the atomic orbital overlap matrix,
\begin{equation}
    S_{\alpha \beta}(R) = \braket{\chi_\alpha | \chi_\beta}
\end{equation}
Automatic differentiation (AD) avoids biases common in numerical methods, such as truncation errors during derivative evaluation. However, the stochastic reconfiguration (SR) procedure in ph-AFQMC introduces a different source of bias in these calculations. SR improves statistical efficiency by periodically eliminating low-weight walkers and replicating high-weight ones. This inherently discontinuous process leads to bias in the derivatives obtained through AD. A comprehensive discussion of this effect is provided in Ref.~\onlinecite{mahajan2023response}, with its impact analyzed further in Sec.~\ref{sec:ch4} pertaining to the current discussion.

In this work, we approximate the PESs of the molecular systems studied in geometry optimization and reaction dynamics using machine learning (ML) potentials. The ML strategies explored in this work are described in Sec.~\ref{sec:ml_efficiency}. Fine-tuning (transfer learning) of foundation models is frequently employed to specialize broadly trained models for specific systems or chemical problems. 
In this approach, the pre-trained parameters of the foundation model are refined using the new data.

KRR is another method considered in this work. KRR models a function $f(x)$ as a weighted sum over kernel functions that measure the similarity between data points.
\begin{equation}
    f(x) = \sum_{i=1}^{n_{train}}\alpha_i K(x,x_i)
\end{equation}
where $x_i$ are the training points, $K$ is a kernel function which quantifies the similarity between two inputs, and $\alpha_i$ are the regression coefficients. These coefficients are obtained by solving,
\begin{equation}
    \alpha = (K + \lambda \mathbb{I})^{-1}y
\end{equation}
where $K$ is a matrix with elements $K_{ij} = K(x_i,x_j)$, $\lambda$ is the regularization parameter, and $y$ represents the training data. 

We use the implementation in sGDML, which trains directly on forces. To do this, the Hessian of the kernel in the Mat\'ern family of kernels is used, which is parameterized by a length parameter. Permutation symmetry of atoms is also considered to make the kernel symmetric. To compare the performance of force-based and energy-based training, we developed an in-house implementation of sGDML that uses the same kernel formalism but is trained exclusively on energies. Further theoretical details are provided in the Supporting Information. 

We also investigated $\Delta$-learning using KRR as an alternative to fine-tuning the NN models. $\Delta$-learning is a technique where the PES of a high-level theory (AFQMC, CCSD(T)) is approximated by starting from a lower-level theory (DFT, HF, or other models). In this framework, the PES from a high-level method such as AFQMC is approximated as
\begin{equation}
V_\mathrm{AFQMC} = \Delta V_{\mathrm{AFQMC - UMA}} + V_\mathrm{UMA},
\end{equation}
where $V_\mathrm{UMA}$ is the baseline PES predicted by the UMA model, and $\Delta V_{\mathrm{AFQMC - UMA}}$ represents the KRR model trained on the corrections.
This method is effective because the energy difference is much smaller and easier to learn, requiring much less data. This effectively requires fewer calculations from the expensive high-level theory.

\section{Results}\label{sec:results}
In this section, we present the results of test calculations in Sec.~\ref{sec:ch4}, geometry optimization in Sec.~\ref{sec:geom_opt}, and machine learning-accelerated nudged elastic band (NEB) calculations in Sec.~\ref{sec:neb}. All the ph-AFQMC calculations used a Hartree-Fock (HF) trial state in the L\"{o}wdin orthonormalized atomic orbitals basis. All HF, DFT calculations and required integrals were obtained using PySCF~\cite{sun2020recent}, and the AFQMC code is available in a public repository~\cite{dqmc_code}. The ph-AFQMC simulations utilized an imaginary timestep of 0.01. All calculations used Dunning's cc-pVXZ basis set, with density-fitted integrals in Cholesky-decomposed form. Automatic differentiation was performed using the JAX package~\cite{jax2018github}. ``uma-s-1p1" checkpoint of UMA was used as the foundation model for all the calculations with the ``omol" task~\cite{levine2025openmolecules2025omol25}.

\subsection{Symmetric C-H bond stretch in methane} \label{sec:ch4}
To evaluate the accuracy of our method, we compare the force $F(R)$ for the symmetric stretch of the C-H bond in methane, as a function of bond distance $R$, obtained using automatic differentiation and finite difference (FD) numerical calculations with the cc-pVDZ basis set. The ph-AFQMC calculation used a UHF trial state. Finite difference calculations were performed using the five-point formula\cite{fdcc} with a step size of 0.1~bohr. Note that FD calculations can be made more efficient using correlated sampling~\cite{Otis25}, however we have not implemented it for this work. FD calculations, of course, have the shortcoming that one must perform a separate calculation for each element of the gradient vector, while rev-AD gives the entire vector in one shot, making it much more computationally efficient. In the current work, we are using FD only for testing purposes, and we overcome the challenge of working with noisy data by using a large step size and higher-order FD formulae. 

\begin{figure}
    \centering
    \includegraphics[width=0.8\linewidth]{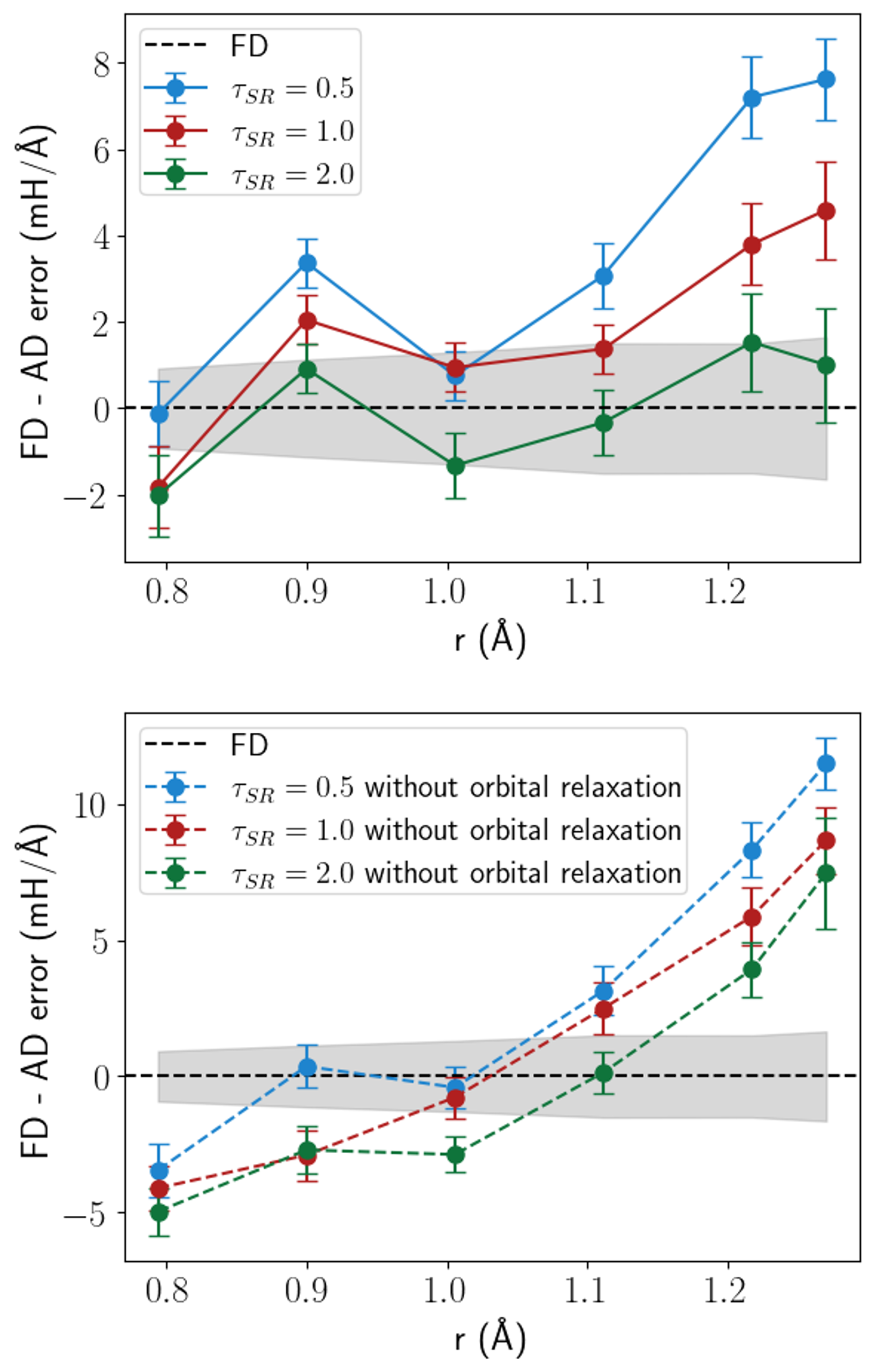}
    \caption{Difference between the nuclear gradients obtained from finite difference (FD) and reverse-mode automatic differentiation (AD) methods for the symmetric stretch of C–H bonds in methane, using different intervals for performing stochastic reconfiguration: (top) with orbital relaxation and (bottom) without orbital relaxation. The grey shaded region represents the stochastic error bars on the FD data.}
    \label{fig:ch4_sr}
\end{figure}

As pointed out in Sec.~\ref{sec:theory}, it is not possible to take the derivative of stochastic reconfiguration (SR) because it is inherently a discontinuous process which nevertheless depends on the input geometry through the imaginary time propagation. Thus, using AD in the presence of SR will lead to biased gradients, and this is one of the limitations of AD. One can reduce the frequency with which the SR is performed to reduce the bias; however doing so leads to larger variance. In addition, to get correct gradients, one has to include the change in energy due to orbital relaxation, as described in Eq.~\ref{eq:gradient}. The ph-AFQMC energy due to orbital relaxation is not automatically included in our code because our code takes the molecular integrals as input and their dependence on the nuclear geometry is calculated separately in PySCF. These terms need to be explicitly included as described in Sec.~\ref{sec:theory}. FIG.~\ref{fig:ch4_sr} shows the impact of both these terms. As evident from the top panel, the gradients converge toward the FD reference as the SR interval is increased, with the $\tau_\mathrm{SR}=2.0$ results always lying within the FD stochastic error bars, albeit with higher noise. Based on this criterion, the $\tau_\mathrm{SR}=1.0$ results effectively control the bias while maintaining acceptable noise levels. It is difficult to prescribe a universally optimal SR interval, which will be system dependent, but for all the remaining calculations we have used $\tau_\mathrm{SR}$ = 1.0.
It is worth noting that in our computational workflow, the gradients (and energies) are fit to an ML model, so the noise and small biases are effectively smoothed.
The bottom panel of FIG.~\ref{fig:ch4_sr} shows that inclusion of orbital relaxation, as described in Eq.~\ref{eq:gradient}, is essential, and omitting this contribution introduces bias in the computed forces.

\subsection{Machine learning models}
\label{sec:ml_efficiency}

\begin{figure*}[]
    \centering
    \includegraphics[width=0.8\linewidth]{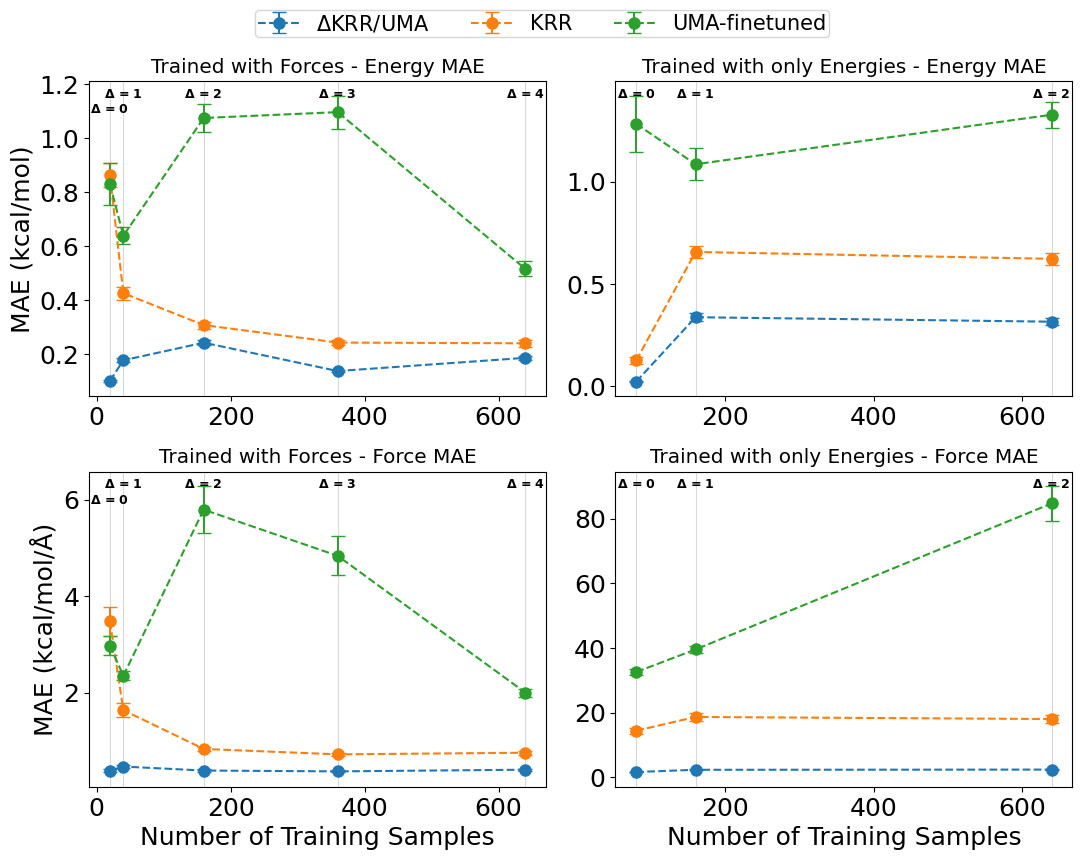}
    \caption{Mean absolute errors (MAEs) of (top) energies (in kcal/mol) and (bottom) forces (in kcal/mol/Å) obtained using different ML methods. The parameter $\Delta$ denotes the standard deviation of random Gaussian noise added to the training dataset, and the x-axis indicates the number of training samples used. The left panels show results for models trained on both energies and forces, and the right panels correspond to models trained exclusively on energies. Note that the training set sizes are scaled with $\Delta$ to approximate a fixed QMC computational budget across datasets (see main text).}
    \label{fig:ml_methods}
\end{figure*}
Using gradients in real-world applications typically requires thousands of evaluations of the underlying electronic structure method, which becomes expensive for computationally demanding methods such as AFQMC. Therefore, it is desirable to develop machine learning (ML) models that can accurately approximate the potential energy surface (PES) while maintaining the accuracy of the underlying method. Moreover, AFQMC energies and gradients inherently contain stochastic error, so it is critical that the ML model learns the physical trends rather than the random fluctuations.

\begin{figure}[]
    \centering
    \includegraphics[width=0.65\linewidth]{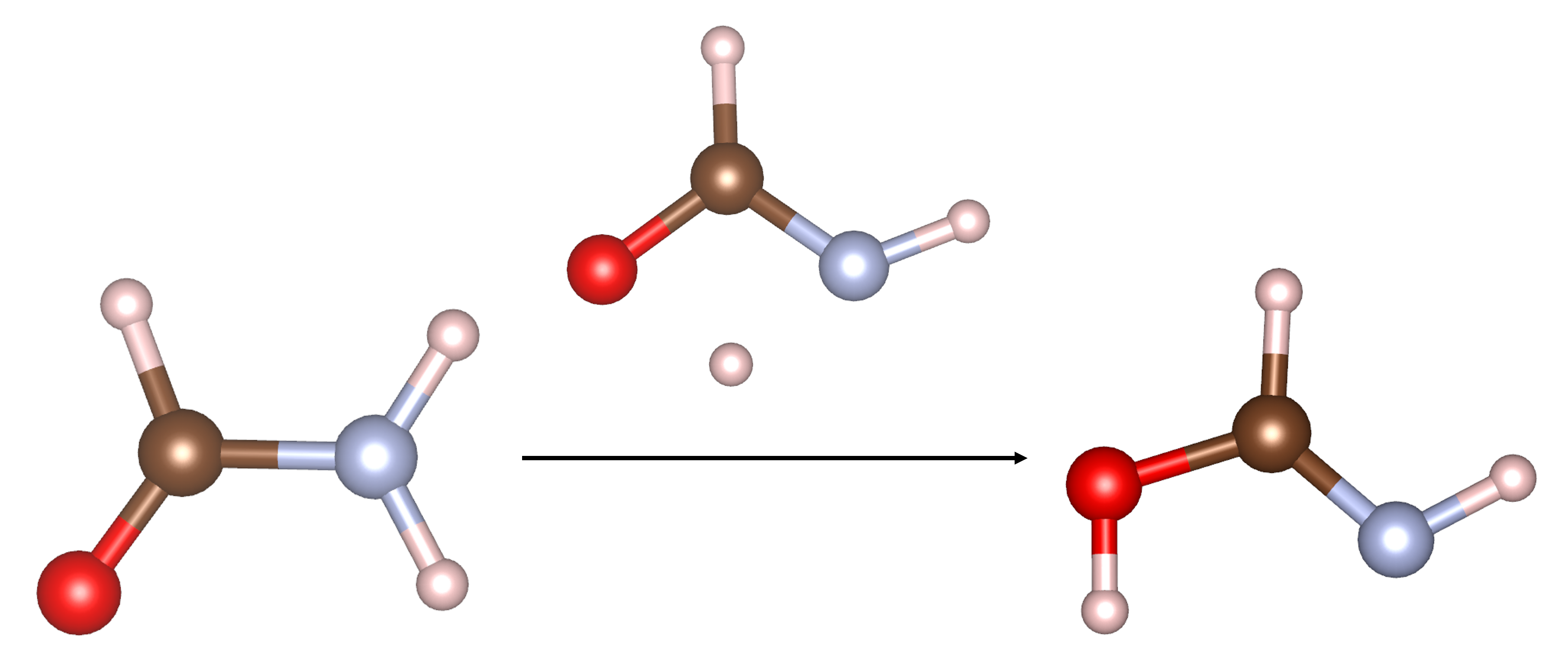}
    \caption{Formamide to formimidic acid hydrogen transfer reaction with a representative transition state shown above the arrow.}
    \label{fig:formamide_formaldehyde}
\end{figure}

The dataset used for testing was generated by performing a molecular dynamics calculation on the product in FIG.~\ref{fig:formamide_formaldehyde} at high temperature (1500~K) using the UMA model. 1000 structures were obtained from this trajectory and their corresponding UCCSD(T) energies and gradients were computed using CFOUR~\cite{matthews2020coupled}. In order to mimic the stochastic error in ph-AFQMC data, we add a random Gaussian noise (with standard deviation $\Delta$) to both energies and forces (in kcal/mol for energy and kcal/mol/\AA~for forces). This also allows us to examine the impact of data noise on model accuracy, as obtaining a small number of highly precise data points from quantum Monte Carlo methods is often computationally harder than obtaining a larger number of noisy samples.
Because the stochastic error in ph-AFQMC scales as $1/\sqrt{N_\text{samples}}$, the number of training points was adjusted accordingly for different noise levels: 
40 points for $\Delta = 1$, 160 points for $\Delta=2$ and proportionally more for higher noise levels (for $\Delta = 0$, we used 20 points).
In this section, we compare (1) kernel ridge regression (KRR), $\Delta$-KRR, and transfer learning; (2) the effect of noise in the training data on the performance of the machine learning models; and (3) the importance of incorporating gradient information, as opposed to using energies alone, during model training.

We first focus on comparing kernel ridge regression (KRR), $\Delta$-KRR, and transfer learning (fine-tuning) the UMA model. Results are compiled in FIG.~\ref{fig:ml_methods}. Neural network models typically require large amounts of data to train an accurate model. However, transfer learning builds on the physics that is already encoded in the foundation model and therefore requires less data to achieve good accuracy. For $\Delta = 0$ and $\Delta = 1$, the fine-tuned UMA model achieves sub-kcal/mol accuracy for energies; however, the corresponding force errors remain relatively large, with mean absolute errors (MAEs) exceeding 2 kcal/mol/\AA. Interestingly, the MAEs for $\Delta = 4$ are smaller than those for $\Delta = 0$, although the performance is still insufficient for practical applications with small datasets.
Next, we considered kernel ridge regression (KRR) using sGDML. KRR is a natural choice for learning from noisy data, as its regularization term effectively smooths out random fluctuations, and it can perform well even with a limited number of training samples. In our tests, KRR consistently outperformed the fine-tuned UMA models. The somewhat larger MAE observed for $\Delta = 0$ case arises because 20 training points are insufficient to train a purely data-driven model from scratch.
Given the favorable performance of KRR, especially for noisy data, we then trained $\Delta$-KRR models with UMA as the baseline. Since UMA evaluations are inexpensive, incorporating them as the lower-level reference in the $\Delta$-learning framework adds negligible overhead. The resulting $\Delta$-KRR/UMA model exhibits excellent accuracy, achieving energy MAEs of approximately 0.2 kcal/mol and force MAEs of around 0.4 kcal/mol/\AA~even for highly noisy data.
These results suggest that reliable models can be trained even in the presence of noisy data, provided that a sufficient amount of training data is available. This robustness can be exploited by performing noisier ph-AFQMC calculations to generate a larger number of training configurations at reduced computational cost.
This is consistent with the conclusions of Ref.~\onlinecite{ceperley2024training}. The specific noise levels and the amount of training data required depend on the problem, and a detailed investigation of these factors is reserved for future work.

We further assessed the importance of force information by training UMA, KRR, and $\Delta$-KRR/UMA models solely on energy data (see FIG.~\ref{fig:ml_methods} (right panel)). Note that even when a model is trained solely on energies, forces can still be obtained analytically from the learned PES. This is important since force evaluation is expensive, and one could generate a larger amount of energy data, especially from ph-AFQMC. Since sGDML only supports training with forces, we implemented an in-house code that uses the same kernel and descriptors as sGDML but adapted for energy-only training. The theoretical details are provided in the Supporting Information. Both fine-tuned UMA and energy-only KRR models perform poorly in predicting forces, while the $\Delta$-KRR/UMA model achieves significantly better but still suboptimal accuracy, with force MAEs in the range of 1.5-2 kcal/mol/\AA. It is evident that forces provide more information, leading to overall better accuracy.
This becomes even more important for larger systems with more degrees of freedom. Including forces provides many more labels per configuration, greatly increasing the information content of the training data and leading to a much better description of the PES. A brief discussion of this effect is provided in the Supplementary Information, where the role of force information is examined for different hydrocarbons.
Based on these observations, $\Delta$-KRR/UMA trained on both energies and forces represents the most promising path forward for ph-AFQMC. For all the following calculations, we use this method.

\subsection{Geometry Optimization}\label{sec:geom_opt}
In order to further test our method, we perform geometry optimization of water and ammonia. The ph-AFQMC gradients were used to train a $\Delta$-learning model, with UMA serving as the lower-level reference. The training geometries for the model were generated via molecular dynamics (MD) simulations using ASE at 1500~K using the UMA model. Running MD at such a high temperature enables sampling of a wide variety of molecular configurations, thereby improving the accuracy of the resulting PES. The same set of geometries was subsequently used for all basis sets. ph-AFQMC/RHF calculations were then performed on these geometries using different basis sets, and the resulting energies and gradients were used to train the KRR model. Details of the hyperparameters employed for training are provided in the Supporting Information. For all calculations, 10 training points were used for the $\Delta$ model.
Geometry optimizations were carried out with geomeTRIC~\cite{geometric} using an interface from PySCF.

The results of these calculations are summarized in TABLE~\ref{tab:geomOpt}. The ph-AFQMC-optimized structures show excellent agreement with those obtained from CCSD(T) across all basis sets considered.
For both H$_2$O and NH$_3$, the deviations in bond lengths and bond angles between ph-AFQMC and CCSD(T) remain consistently small, with differences of less than 0.001\AA ~in bond lengths and 0.1\degree ~in bond angles.
In comparison, B3LYP exhibits average deviations of 0.003\AA ~and 0.002\AA ~in bond lengths, and 0.6\degree ~and 0.5\degree ~in bond angles, for H$_2$O and NH$_3$, respectively. These results also demonstrate the absence of hysteresis from the UMA reference, in the sense that the $\Delta$-learning model does not remain restricted to the level at which the UMA model was trained, but instead achieves consistent and accurate results across different basis sets, even with a small AFQMC training set. The agreement with high-level theory like CCSD(T) is also encouraging, allowing the use of AFQMC gradients to solve more challenging problems.

\begin{table*}[]
\caption{Optimized geometries obtained using ph-AFQMC compared to those of CCSD(T) with different basis sets. Experimental results are obtained from Ref.~\onlinecite{johnson1999nist}.}
\label{tab:geomOpt}
\begin{tabular}{cccccc}
\hline
 &
   &
  \multicolumn{2}{c}{\ce{H2O}} &
  \multicolumn{2}{c}{\ce{NH3}} \\
\multirow{-2}{*}{Method} & \multirow{-2}{*}{Basis} & Bond length (\AA) & Bond angle & Bond length (\AA) & Bond angle \\ \hline
B3LYP &
   &
  0.969 &
  102.73 &
  1.025 &
  104.32 \\
ML - AFQMC &
   &
  0.966 &
  101.89 &
  \cellcolor[HTML]{FFFFFF}1.026 &
  \cellcolor[HTML]{FFFFFF}103.38 \\
CCSD(T) &
  \multirow{-3}{*}{cc-pVDZ} &
  0.966 &
  101.96 &
  1.026 &
  103.59 \\ \hline
B3LYP &
   &
  0.961 &
  104.53 &
  1.014 &
  106.50 \\
ML - AFQMC &
   &
  0.957 &
  103.76 &
  1.010 &
  105.99 \\
CCSD(T) &
  \multirow{-3}{*}{cc-pVTZ} &
  0.958 &
  103.73 &
  1.011 &
  106.06 \\ \hline
B3LYP &
   &
  0.960 &
  104.88 &
  1.013 &
  106.86 \\
ML - AFQMC &
   &
  \cellcolor[HTML]{FFFFFF}0.956 &
  \cellcolor[HTML]{FFFFFF}104.34 &
  1.009 &
  106.45 \\
CCSD(T) &
  \multirow{-3}{*}{cc-pvQZ} &
  0.956 &
  104.25 &
  1.010 &
  106.36 \\ \hline
\multicolumn{1}{l}{Experiment} &
  \multicolumn{1}{l}{} &
  0.958 &
  104.50 &
  1.012 &
  106.67 \\ \hline
\end{tabular}
\end{table*}

\subsection{Nudged elastic band}\label{sec:neb}
In this section, we present the results of a nudged elastic band (NEB) calculation to identify the transition state of the proton transfer reaction between formamide and formimidic acid. Proton transfer reactions of this type are biologically significant and are often used as model systems for investigating tautomerization processes in nucleic acids~\cite{liang2004proton}. 

The reactant and product geometries used in the NEB calculation were optimized at the B3LYP/cc-pVTZ level, with the optimized structures provided in the Supplementary Information. In order to construct the training dataset, we first performed an NEB calculation with UMA. All the geometries that were sampled during this calculation were saved, and a subset of 100 structures was then selected as the training dataset. Performing the initial NEB calculation with a foundation model provides an intuitive and efficient means of selecting training points for the computationally expensive AFQMC calculations, thereby avoiding reliance on arbitrary geometries.
To choose the best hyperparameters, a subset of 60 geometries from the training dataset was first used to train models over a grid of length scale parameters, $\sigma \in (1,30)$ and regularization parameters, $\lambda \in (1,10^{-14})$. The remaining geometries were used as a testing set to find the combination of $\sigma$ and $\lambda$ yielding the lowest error. This pair of hyperparameters was used to train the subsequent models. We used a combination of ($\sigma$,$\lambda$) = (23,$10^{-8}$) for training the $\Delta$ model. Notably, we observed that multiple combinations of hyperparameters gave rise to similar results, indicating some degree of robustness in choosing the hyperparameters. The NEB calculations were conducted with the Atomic Simulation Environment (ASE), with a maximum force convergence criterion (fmax) of 0.05 eV/$\AA$. To estimate the stochastic error bars for the NEB calculations, 100 independent NEB calculations were performed using models trained on datasets in which random Gaussian noise was added to the training points. 
The standard deviation of this noise was chosen to be the stochastic error in the original training data. This procedure provides a practical means of propagating the intrinsic stochastic uncertainty of the ph-AFQMC training data into the NEB results.

\begin{figure}
    \centering
    \includegraphics[width=0.75\linewidth]{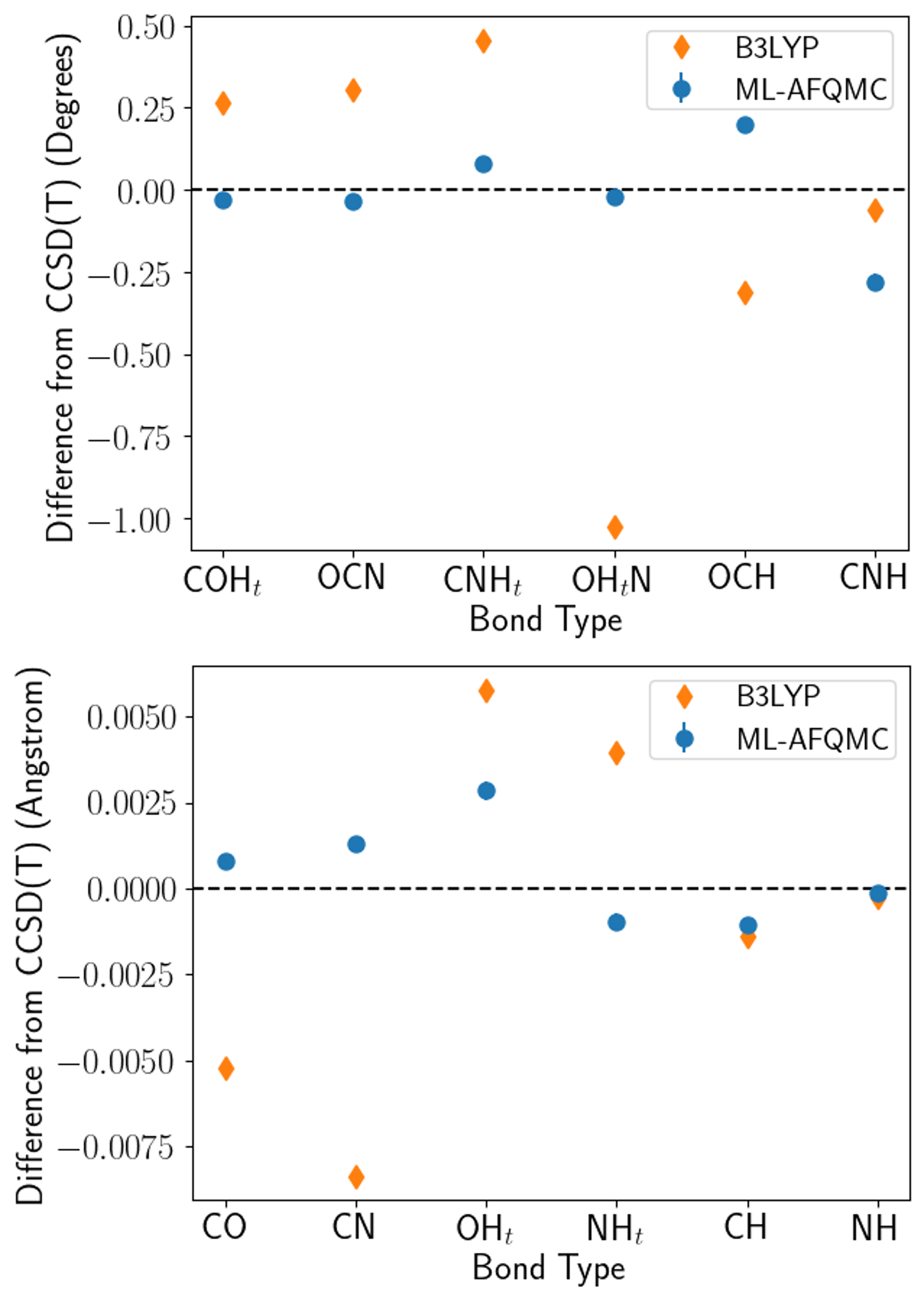}
    \caption{Results of the NEB calculation for the formamide–formimidic acid tautomerization reaction. Differences in bond lengths (top) and bond angles (bottom) for all bonds in the transition state, as predicted by DFT (B3LYP) and AFQMC, relative to CCSD(T).}
    \label{fig:formamide_bl_ba}
\end{figure}

From the structural parameters (see FIG.~\ref{fig:formamide_bl_ba}), it is evident that the transition state (TS) obtained from AFQMC closely resembles the structure obtained using CCSD(T) and B3LYP, with AFQMC performing better than B3LYP. The average bond length difference between AFQMC and CCSD(T) is just 0.001 Å, to be compared to the 0.004 Å difference for B3LYP. Similarly, the average bond angle differences are 0.1\degree ~and 0.4\degree ~for AFQMC and B3LYP, respectively, compared to CCSD(T).
In terms of barrier heights, AFQMC predicted a forward barrier $(\Delta E_f^\dagger)$ of 45.23(4) kcal/mol, which is similar to what was predicted by B3LYP (44.73 kcal/mol) and CCSD(T) (45.66 kcal/mol). The Reverse barrier height ($\Delta E_r^\dagger$) shows somewhat larger deviations, where AFQMC predicted 33.31(4) kcal/mol, whereas B3LYP and CCSD(T) predicted 32.28 kcal/mol and 34.94 kcal/mol, respectively. We further performed AFQMC energy calculations with UCISD trial wavefunction on a random subset of 10 ML-predicted TS structures and the endpoints, which corrected the barrier heights to $\Delta E_f^\dagger$ = 45.0(1) and $\Delta E_r^\dagger$ = 34.5(1).

\begin{table}[h!]
\caption{Barrier heights of  formamide to formimidic acid reaction in kcal/mol. CCSD(T) transition state was obtained by performing an NEB calculation with $\Delta$-KRR/UMA method with 200 UCCSD(T) training points.}
\label{tab:formamide}
\begin{tabular}{ccc}
Method     & $\Delta E^\dagger_f$ & $\Delta E^\dagger_r$ \\ \hline
B3LYP      & 44.73              & 32.28              \\
ML-AFQMC   & 45.23(4)           & 33.31(4)           \\
AFQMC/UCISD & 45.0(1)            & 34.5(1)            \\
CCSD(T)    & 45.66              & 34.94              \\ \hline
\end{tabular}
\end{table}

\subsection{Scaling}
\begin{figure}[]
    \centering
    \includegraphics[width=0.75\linewidth]{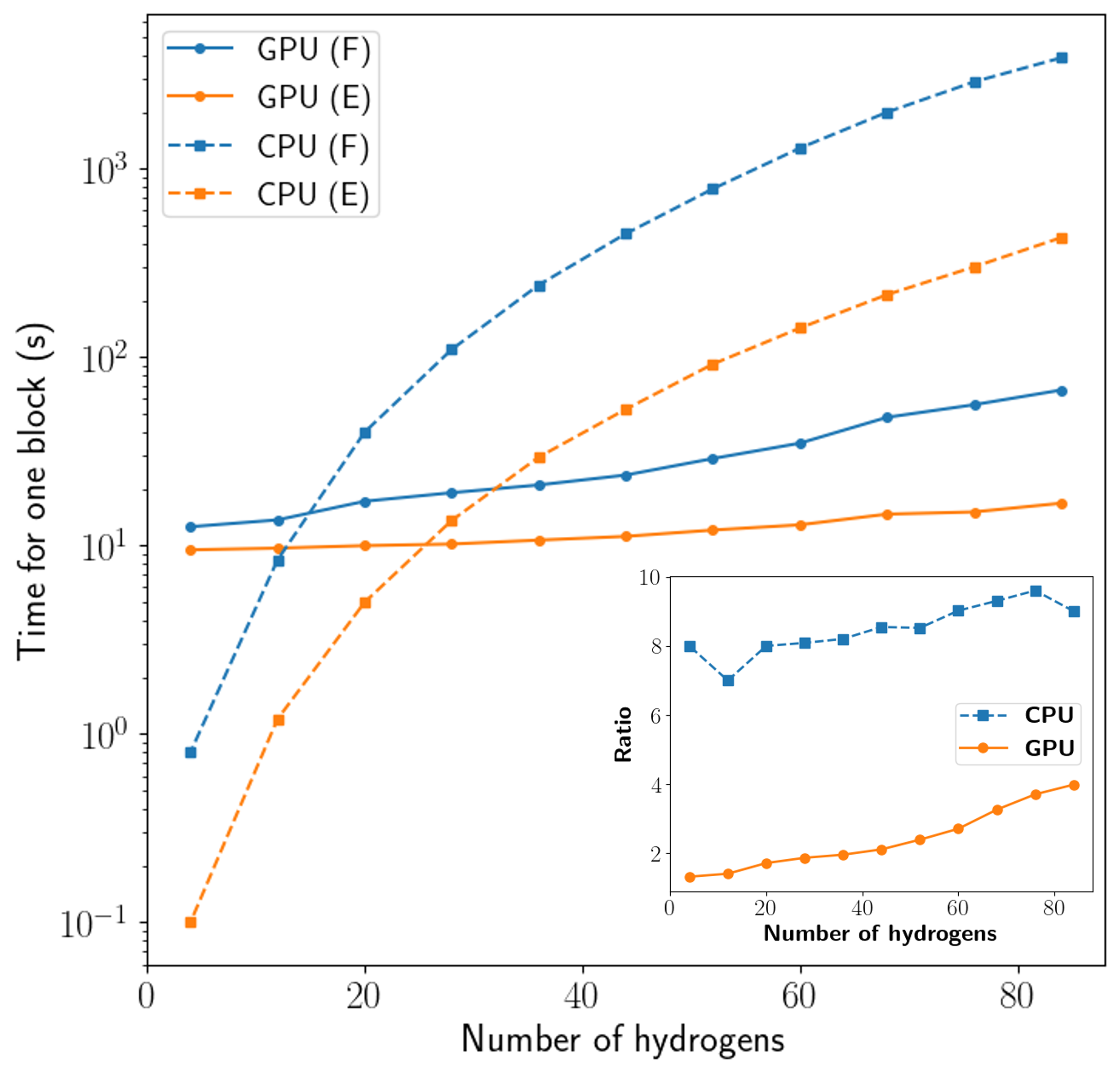} 
    \caption{Scaling of AFQMC energy and rev-AD force evaluations for a 1-D hydrogen chain in the STO-6G basis. Wall-clock timings versus system size are shown for energy-only (orange) and force (blue) calculations, comparing CPU (dashed) and GPU (solid) performance for identical sampling effort. The inset shows the ratio of the wall-clock time for force evaluations to that for energy-only evaluations.}
    \label{fig:scaling}
\end{figure}
To estimate the scaling of the algorithm, we performed test calculations on a 1-D hydrogen chain in the STO-6G basis set, with the bond length between adjacent hydrogen atoms set to 1.6 a.u. The results of these benchmark calculations are shown in FIG.~\ref{fig:scaling}. GPU calculations were performed on a single NVIDIA A100 GPU, and CPU calculations were performed on a 20-core Intel Xeon processor. The total number of walkers was set to 200, and in the CPU calculation, this was distributed over the 20 cores. All other parameters were kept constant to ensure a fair comparison. As expected, the evaluation of forces using rev-AD is only a constant prefactor more expensive than computing energies alone. This overhead corresponds to a factor of 8-10 on CPU and 2-4 for GPU calculations. The largest system studied had 84 hydrogens before the GPU ran out of memory. We also note that one-to-one comparison of absolute timings is ill-advised, as performance is heavily dependent on the system architecture. Nevertheless, these results indicate that GPUs generally provide speedup over CPU calculations (except for very small systems). Current memory issues in GPUs can be addressed by parallelizing the code across multiple GPUs, which will be explored in future work.

\section{Conclusion}
\label{sec:conc}
In this work, we presented an algorithm to obtain nuclear gradients in ph-AFQMC using automatic differentiation, with the goal of achieving efficient and accurate gradient evaluations. Comparison with finite difference results for the symmetric stretch of CH$_4$ demonstrated that the proposed method achieves excellent accuracy, despite the bias introduced by the stochastic reconfiguration procedure. The cost of gradient evaluation was found to be approximately 2–4 times that of an energy calculation on GPUs and 8–10 times on CPUs.

We further explored different machine learning (ML) models to approximate AFQMC potential energy surfaces. Among these, the $\Delta$-KRR model using UMA as the baseline was found to be the most accurate and robust in handling noisy ph-AFQMC data. It was also found that training ML models with just energy data resulted in large errors in gradient predictions, and the relative advantage of including gradient training data increased with larger system sizes. How these trends translate to large-scale discovery and long-time dynamics remains an open question. Geometry optimization and nudged elastic band (NEB) calculations performed using these ML models trained on ph-AFQMC gradients show excellent agreement with CCSD(T) results showing its accuracy. 

Reverse-mode AD introduces a large memory overhead, which requires careful optimization of the implementation to efficiently run on GPUs with limited memory. Incorporating efficient checkpointing schemes and enabling parallelization across multiple GPUs can help in this regard. These improvements, along with applications to complex systems and frameworks such as QM/MM, will be explored in the future. We believe that the developments presented here pave the way for highly accurate and scalable molecular dynamics and reaction dynamics simulations.

\section*{ACKNOWLEDGMENTS}
S.S. and J.S.K were partially supported by NSF CHE-
2145209. A.M. thanks David Reichman for support. This work utilized the Blanca condo computing resource at the University of Colorado Boulder. Blanca is jointly funded by computing users and the University of Colorado Boulder. This work used Delta at National Center for Supercomputing Applications through allocation CHE240172 from the Advanced Cyberinfrastructure Coordination Ecosystem: Services \& Support (ACCESS) program, which is supported by U.S. National Science Foundation grants \#2138259, \#2138286, \#2138307, \#2137603, and \#2138296.

\section*{Data availability}
The code used for obtaining the gradients is available in a public GitHub repository at Ref. \citenum{dqmc_code}.

\section*{AUTHOR INFORMATION}
\noindent\textbf{Jo S. Kurian} -- Department of Chemistry, University of Colorado, Boulder, CO 80302, USA;\\ \url{https://orcid.org/0000-0001-9273-1043};\\ Email: \url{jokurian12@gmail.com}\\

\noindent\textbf{Ankit Mahajan} -- Department of Chemistry, Columbia University, New York, NY 10027, USA;\\
\url{https://orcid.org/0000-0002-2138-3798}\\ \\
\textbf{Sandeep Sharma} -- Department of Chemistry, University of Colorado, Boulder, CO 80302, USA; Division of Chemistry and Chemical Engineering, California Institute of Technology, Pasadena, California 91125, USA; Marcus Center for Theoretical Chemistry, Pasadena CA 91125, USA;
\url{https://orcid.org/0000-0002-6598-8887};\\Email: \url{sanshar@gmail.com}\\ \\


\subsection*{Notes}
The authors declare no competing financial interest.


\bibliographystyle{achemso}

\bibliography{references}

@string{JCTC = "J. Chem. Theory Comput."}

@article{Otis25,
author = {Otis, Leon and Gudivada, Saisrinivas and Friede, Marvin and Shee, James},
title = {A Scalable Route to First-Order Response Properties with Correlated Sampling Phaseless Auxiliary-Field Quantum Monte Carlo},
journal = {The Journal of Physical Chemistry Letters},
volume = {16},
number = {44},
pages = {11390-11397},
year = {2025},
}

@article{mahajan2023response,
  title={Response properties in phaseless auxiliary field quantum Monte Carlo},
  author={Mahajan, Ankit and Kurian, Jo S and Lee, Joonho and Reichman, David R and Sharma, Sandeep},
  journal={The Journal of Chemical Physics},
  volume={159},
  number={18},
  year={2023},
  publisher={AIP Publishing}
}

@misc{fdcc,
  title={Finite Difference Coefficients Calculator},
  author={Taylor, Cameron R.},
  year={2016},
  howpublished="\url{https://web.media.mit.edu/~crtaylor/calculator.html}"
}

@article{hubbardBenchmark2015,
	title        = {Solutions of the Two-Dimensional Hubbard Model: Benchmarks and Results from a Wide Range of Numerical Algorithms},
	author = {LeBlanc, J. P. F. and Antipov, Andrey E. and Becca, Federico and Bulik, Ireneusz W. and Chan, Garnet Kin-Lic and Chung, Chia-Min and Deng, Youjin and Ferrero, Michel and Henderson, Thomas M. and Jim\'enez-Hoyos, Carlos A. and others},
	year         = 2015,
	month        = {Dec},
	journal      = {Phys. Rev. X},
	publisher    = {American Physical Society},
	volume       = 5,
	pages        = {041041},
	doi          = {10.1103/PhysRevX.5.041041},
	collaboration = {Simons Collaboration on the Many-Electron Problem},
	issue        = 4,
	numpages     = 28
}

@article{hydrogenBenchmark2017,
	title        = {Towards the Solution of the Many-Electron Problem in Real Materials: Equation of State of the Hydrogen Chain with State-of-the-Art Many-Body Methods},
	author       = {Motta, Mario and Ceperley, David M. and Chan, Garnet Kin-Lic and Gomez, John A. and Gull, Emanuel and Guo, Sheng and Jim\'enez-Hoyos, Carlos A. and Lan, Tran Nguyen and Li, Jia and Ma, Fengjie and others},
	year         = 2017,
	month        = {Sep},
	journal      = {Phys. Rev. X},
	publisher    = {American Physical Society},
	volume       = 7,
	pages        = {031059},
	doi          = {10.1103/PhysRevX.7.031059},
	collaboration = {Simons Collaboration on the Many-Electron Problem},
	issue        = 3,
	numpages     = 28
}

@article{transitionMetalOxides2020,
	title        = {Direct Comparison of Many-Body Methods for Realistic Electronic Hamiltonians},
	author       = {Williams, Kiel T. and Yao, Yuan and Li, Jia and Chen, Li and Shi, Hao and Motta, Mario and Niu, Chunyao and Ray, Ushnish and Guo, Sheng and Anderson, Robert J. and others},
	year         = 2020,
	month        = {Feb},
	journal      = {Phys. Rev. X},
	publisher    = {American Physical Society},
	volume       = 10,
	pages        = {011041},
	doi          = {10.1103/PhysRevX.10.011041},
	collaboration = {Simons Collaboration on the Many-Electron Problem},
	issue        = 1,
	numpages     = 9
}

@article{Shee2020afqmc,
	title        = {Predicting Ligand-Dissociation Energies of 3d Coordination Complexes with Auxiliary-Field Quantum Monte Carlo},
	author       = {Rudshteyn, B. and Coskun, D. and Weber, J. L. and Arthur, E. J. and Zhang, S. and Reichman, D. R. and Friesner, R. A. and Shee, J.},
	year         = 2020,
	journal      = {J. Chem. Theory Comput.},
	volume       = 16,
	pages        = 3041,
    doi = {10.1021/acs.jctc.0c00070}
}

@article{Friesner2022afqmc_benchmark,
	title        = {Calculation of Metallocene Ionization Potentials via Auxiliary Field Quantum Monte Carlo: Toward Benchmark Quantum Chemistry for Transition Metals},
	author       = {Rudshteyn, B. and Weber, J. L. and Coskun, D. and Devlaminck, P. A. and Zhang, S. and Reichman, D. R. and Shee, J. and Friesner, R. A.},
	year         = 2022,
	journal      = {J. Chem. Theory Comput.},
	volume       = 18,
	pages        = 2845,
    doi = {10.1021/acs.jctc.1c01071}
}

@article{Lee2020FullereneFePorphyrin,
	title        = {Utilizing Essential Symmetry Breaking in Auxiliary-Field Quantum Monte Carlo: Application to the Spin Gaps of the C36 Fullerene and an Iron Porphyrin Model Complex},
	author       = {Lee, Joonho and Malone, Fionn D. and Morales, Miguel A.},
	year         = 2020,
	journal      = JCTC,
	volume       = 16,
	number       = 5,
	pages        = {3019--3027},
	doi          = {10.1021/acs.jctc.0c00055}
}

@article{sukurma2023benchmark,
	title        = {Benchmark Phaseless Auxiliary-Field Quantum Monte Carlo Method for Small Molecules},
	author       = {Sukurma, Zoran and Schlipf, Martin and Humer, Moritz and Taheridehkordi, Amir and Kresse, Georg},
	year         = 2023,
	journal      = JCTC,
	volume       = 19,
	number       = 15,
	pages        = {4921--4934},
	doi          = {10.1021/acs.jctc.3c00322}
}

@article{Sharma2022afqmc,
	title        = {Selected configuration interaction wave functions in phaseless auxiliary field quantum Monte Carlo},
	author       = {Mahajan, A. and Lee, J. and Sharma, S.},
	year         = 2022,
	journal      = {J. Chem. Phys.},
	volume       = 156,
	pages        = 174111,
doi = {10.1063/5.0087047}
}

@article{eskridge2019local,
  title={Local embedding and effective downfolding in the auxiliary-field quantum Monte Carlo method},
  author={Eskridge, Brandon and Krakauer, Henry and Zhang, Shiwei},
  journal={Journal of Chemical Theory and Computation},
  volume={15},
  number={7},
  pages={3949--3959},
  year={2019},
  publisher={ACS Publications}
}

@article{assaraf2003zero,
  title={Zero-variance zero-bias principle for observables in quantum Monte Carlo: Application to forces},
  author={Assaraf, Roland and Caffarel, Michel},
  journal={The Journal of Chemical Physics},
  volume={119},
  number={20},
  pages={10536--10552},
  year={2003},
  publisher={American Institute of Physics}
}

@article{badinski2010methods,
  title={Methods for calculating forces within quantum Monte Carlo simulations},
  author={Badinski, A and Haynes, Peter D and Trail, JR and Needs, RJ},
  journal={Journal of Physics: Condensed Matter},
  volume={22},
  number={7},
  pages={074202},
  year={2010},
  publisher={IOP Publishing}
}

@article{filippi2000correlated,
  title={Correlated sampling in quantum Monte Carlo: A route to forces},
  author={Filippi, Claudia and Umrigar, CJ},
  journal={Physical Review B},
  volume={61},
  number={24},
  pages={R16291},
  year={2000},
  publisher={APS}
}

@article{attaccalite2008stable,
  title={Stable liquid hydrogen at high pressure by a novel ab initio molecular-dynamics calculation},
  author={Attaccalite, Claudio and Sorella, Sandro},
  journal={Physical review letters},
  volume={100},
  number={11},
  pages={114501},
  year={2008},
  publisher={APS}
}

@article{moroni2014practical,
  title={Practical schemes for accurate forces in quantum Monte Carlo},
  author={Moroni, S and Saccani, S and Filippi, Claudia},
  journal={Journal of chemical theory and computation},
  volume={10},
  number={11},
  pages={4823--4829},
  year={2014},
  publisher={ACS Publications}
}

@article{pathak2020light,
  title={A light weight regularization for wave function parameter gradients in quantum Monte Carlo},
  author={Pathak, Shivesh and Wagner, Lucas K},
  journal={AIP Advances},
  volume={10},
  number={8},
  year={2020},
  publisher={AIP Publishing}
}

@article{motta2018communication,
  title={Communication: Calculation of interatomic forces and optimization of molecular geometry with auxiliary-field quantum Monte Carlo},
  author={Motta, Mario and Zhang, Shiwei},
  journal={The Journal of Chemical Physics},
  volume={148},
  number={18},
  year={2018},
  publisher={AIP Publishing}
}

@article{jiang2022general,
  title={General Analytical Nuclear Forces and Molecular Potential Energy Surface from Full Configuration Interaction Quantum Monte Carlo},
  author={Jiang, Tonghuan and Fang, Wei and Alavi, Ali and Chen, Ji},
  journal={Journal of Chemical Theory and Computation},
  volume={18},
  number={12},
  pages={7233--7242},
  year={2022},
  publisher={ACS Publications}
}

@article{ceperley2024training,
  title={Training models using forces computed by stochastic electronic structure methods},
  author={Ceperley, David M and Jensen, Scott and Yang, Yubo and Niu, Hongwei and Pierleoni, Carlo and Holzmann, Markus},
  journal={Electronic Structure},
  volume={6},
  number={1},
  pages={015011},
  year={2024},
  publisher={IOP Publishing}
}

@article{slootman2024accurate,
  title={Accurate quantum Monte Carlo forces for machine-learned force fields: Ethanol as a benchmark},
  author={Slootman, Emiel and Poltavsky, Igor and Shinde, Ravindra and Cocomello, Jacopo and Moroni, Saverio and Tkatchenko, Alexandre and Filippi, Claudia},
  journal={Journal of chemical theory and computation},
  volume={20},
  number={14},
  pages={6020--6027},
  year={2024},
  publisher={ACS Publications}
}

@article{chen2023computation,
  title={Computation of forces and stresses in solids: Towards accurate structural optimization with auxiliary-field quantum Monte Carlo},
  author={Chen, Siyuan and Zhang, Shiwei},
  journal={Physical Review B},
  volume={107},
  number={19},
  pages={195150},
  year={2023},
  publisher={APS}
}

@article{sorella2010algorithmic,
  title={Algorithmic differentiation and the calculation of forces by quantum Monte Carlo},
  author={Sorella, Sandro and Capriotti, Luca},
  journal={The Journal of chemical physics},
  volume={133},
  number={23},
  year={2010},
  publisher={AIP Publishing}
}

@article{zhang2023automatic,
  title={Automatic differentiable Monte Carlo: theory and application},
  author={Zhang, Shi-Xin and Wan, Zhou-Quan and Yao, Hong},
  journal={Physical Review Research},
  volume={5},
  number={3},
  pages={033041},
  year={2023},
  publisher={APS}
}

@phdthesis{poole2014calculating,
  title={Calculating derivatives within quantum Monte Carlo},
  author={Poole, Thomas},
  year={2014},
  school={Imperial College London}
}

@article{motta2018ab,
  title={Ab initio computations of molecular systems by the auxiliary-field quantum Monte Carlo method},
  author={Motta, Mario and Zhang, Shiwei},
  journal={Wiley Interdisciplinary Reviews: Computational Molecular Science},
  volume={8},
  number={5},
  pages={e1364},
  year={2018},
  publisher={Wiley Online Library}
}

@article{chmiela2017machine,
  title={Machine learning of accurate energy-conserving molecular force fields},
  author={Chmiela, Stefan and Tkatchenko, Alexandre and Sauceda, Huziel E and Poltavsky, Igor and Sch{\"u}tt, Kristof T and M{\"u}ller, Klaus-Robert},
  journal={Science advances},
  volume={3},
  number={5},
  pages={e1603015},
  year={2017},
  publisher={American Association for the Advancement of Science}
}

@article{chmiela2018towards,
  title={Towards exact molecular dynamics simulations with machine-learned force fields},
  author={Chmiela, Stefan and Sauceda, Huziel E and M{\"u}ller, Klaus-Robert and Tkatchenko, Alexandre},
  journal={Nature communications},
  volume={9},
  number={1},
  pages={3887},
  year={2018},
  publisher={Nature Publishing Group UK London}
}

@misc{wood2025umafamilyuniversalmodels,
      title={UMA: A Family of Universal Models for Atoms}, 
      author={Brandon M. Wood and Misko Dzamba and Xiang Fu and Meng Gao and Muhammed Shuaibi and Luis Barroso-Luque and Kareem Abdelmaqsoud and Vahe Gharakhanyan and John R. Kitchin and Daniel S. Levine and Kyle Michel and Anuroop Sriram and Taco Cohen and Abhishek Das and Ammar Rizvi and Sushree Jagriti Sahoo and Zachary W. Ulissi and C. Lawrence Zitnick},
      year={2025},
      eprint={2506.23971},
      archivePrefix={arXiv},
      primaryClass={cs.LG},
      url={https://arxiv.org/abs/2506.23971}, 
}

@software{jax2018github,
  author = {James Bradbury and Roy Frostig and Peter Hawkins and Matthew James Johnson and Chris Leary and Dougal Maclaurin and George Necula and Adam Paszke and Jake Vander{P}las and Skye Wanderman-{M}ilne and Qiao Zhang},
  title = {{JAX}: composable transformations of {P}ython+{N}um{P}y programs},
  url = {http://github.com/jax-ml/jax},
  version = {0.3.13},
  year = {2018},
}

@misc{levine2025openmolecules2025omol25,
      title={The Open Molecules 2025 (OMol25) Dataset, Evaluations, and Models}, 
      author={Daniel S. Levine and Muhammed Shuaibi and Evan Walter Clark Spotte-Smith and Michael G. Taylor and Muhammad R. Hasyim and Kyle Michel and Ilyes Batatia and Gábor Csányi and Misko Dzamba and Peter Eastman and Nathan C. Frey and Xiang Fu and Vahe Gharakhanyan and Aditi S. Krishnapriyan and Joshua A. Rackers and Sanjeev Raja and Ammar Rizvi and Andrew S. Rosen and Zachary Ulissi and Santiago Vargas and C. Lawrence Zitnick and Samuel M. Blau and Brandon M. Wood},
      year={2025},
      eprint={2505.08762},
      archivePrefix={arXiv},
      primaryClass={physics.chem-ph},
      url={https://arxiv.org/abs/2505.08762}, 
}

@article{geometric,
    author = {Wang, Lee-Ping and Song, Chenchen},
    title = {Geometry optimization made simple with translation and rotation coordinates},
    journal = {The Journal of Chemical Physics},
    volume = {144},
    number = {21},
    pages = {214108},
    year = {2016},
    month = {06},
    issn = {0021-9606},
    doi = {10.1063/1.4952956},
    url = {https://doi.org/10.1063/1.4952956},
    eprint = {https://pubs.aip.org/aip/jcp/article-pdf/doi/10.1063/1.4952956/15512057/214108\_1\_online.pdf},
}

@article{liang2004proton,
  title={Proton Transfer of Formamide+ n H2O (n= 0- 3): Protective and Assistant Effect of the Water Molecule},
  author={Liang, Wanchun and Li, Haoran and Hu, Xingbang and Han, Shijun},
  journal={The Journal of Physical Chemistry A},
  volume={108},
  number={46},
  pages={10219--10224},
  year={2004},
  publisher={ACS Publications}
}

@article{tirelli2022high,
  title={High-pressure hydrogen by machine learning and quantum Monte Carlo},
  author={Tirelli, Andrea and Tenti, Giacomo and Nakano, Kousuke and Sorella, Sandro},
  journal={Physical Review B},
  volume={106},
  number={4},
  pages={L041105},
  year={2022},
  publisher={APS}
}

@article{huang2022machine,
  title={Machine learning diffusion Monte Carlo forces},
  author={Huang, Cancan and Rubenstein, Brenda M},
  journal={The Journal of Physical Chemistry A},
  volume={127},
  number={1},
  pages={339--355},
  year={2022},
  publisher={ACS Publications}
}

@article{huang2023toward,
  title={Toward DMC accuracy across chemical space with scalable $\Delta$-QML},
  author={Huang, Bing and von Lilienfeld, O Anatole and Krogel, Jaron T and Benali, Anouar},
  journal={Journal of Chemical Theory and Computation},
  volume={19},
  number={6},
  pages={1711--1721},
  year={2023},
  publisher={ACS Publications}
}

@article{sun2020recent,
  title={Recent developments in the PySCF program package},
  author={Sun, Qiming and Zhang, Xing and Banerjee, Samragni and Bao, Peng and Barbry, Marc and Blunt, Nick S and Bogdanov, Nikolay A and Booth, George H and Chen, Jia and Cui, Zhi-Hao and others},
  journal={The Journal of chemical physics},
  volume={153},
  number={2},
  year={2020},
  publisher={AIP Publishing}
}

@misc{dqmc_code,
  howpublished = "\url{https://github.com/ankit76/ad_afqmc/}"
}

@article{matthews2020coupled,
  title={Coupled-cluster techniques for computational chemistry: The CFOUR program package},
  author={Matthews, Devin A and Cheng, Lan and Harding, Michael E and Lipparini, Filippo and Stopkowicz, Stella and Jagau, Thomas-C and Szalay, P{\'e}ter G and Gauss, J{\"u}rgen and Stanton, John F},
  journal={The Journal of Chemical Physics},
  volume={152},
  number={21},
  year={2020},
  publisher={AIP Publishing}
}

@article{johnson1999nist,
  title={NIST 101. Computational chemistry comparison and benchmark database},
  author={Johnson III, Russell D},
  year={1999},
  publisher={Russell D. Johnson III}
}

@article{car1985unified,
  title={Unified approach for molecular dynamics and density-functional theory},
  author={Car, Richard and Parrinello, Mark},
  journal={Physical review letters},
  volume={55},
  number={22},
  pages={2471},
  year={1985},
  publisher={APS}
}

@article{pulay2014analytical,
  title={Analytical derivatives, forces, force constants, molecular geometries, and related response properties in electronic structure theory},
  author={Pulay, Peter},
  journal={Wiley Interdisciplinary Reviews: Computational Molecular Science},
  volume={4},
  number={3},
  pages={169--181},
  year={2014},
  publisher={Wiley Online Library}
}

@article{eckert1997ab,
  title={Ab initio geometry optimization for large molecules},
  author={Eckert, Frank and Pulay, Peter and Werner, Hans-Joachim},
  journal={Journal of computational chemistry},
  volume={18},
  number={12},
  pages={1473--1483},
  year={1997},
  publisher={Wiley Online Library}
}

@article{pulay1969ab,
  title={Ab initio calculation of force constants and equilibrium geometries in polyatomic molecules: I. Theory},
  author={Pulay, Peter},
  journal={Molecular Physics},
  volume={17},
  number={2},
  pages={197--204},
  year={1969},
  publisher={Taylor \& Francis}
}

@article{reveles2004geometry,
  title={Geometry optimization in density functional methods},
  author={Reveles, J Ulises and K{\"o}ster, Andreas M},
  journal={Journal of computational chemistry},
  volume={25},
  number={9},
  pages={1109--1116},
  year={2004},
  publisher={Wiley Online Library}
}

@article{unke2021machine,
  title={Machine learning force fields},
  author={Unke, Oliver T and Chmiela, Stefan and Sauceda, Huziel E and Gastegger, Michael and Poltavsky, Igor and Schutt, Kristof T and Tkatchenko, Alexandre and Muller, Klaus-Robert},
  journal={Chemical Reviews},
  volume={121},
  number={16},
  pages={10142--10186},
  year={2021},
  publisher={ACS Publications}
}

@article{batzner20223,
  title={E (3)-equivariant graph neural networks for data-efficient and accurate interatomic potentials},
  author={Batzner, Simon and Musaelian, Albert and Sun, Lixin and Geiger, Mario and Mailoa, Jonathan P and Kornbluth, Mordechai and Molinari, Nicola and Smidt, Tess E and Kozinsky, Boris},
  journal={Nature communications},
  volume={13},
  number={1},
  pages={2453},
  year={2022},
  publisher={Nature Publishing Group UK London}
}

@article{qiao2020orbnet,
  title={OrbNet: Deep learning for quantum chemistry using symmetry-adapted atomic-orbital features},
  author={Qiao, Zhuoran and Welborn, Matthew and Anandkumar, Animashree and Manby, Frederick R and Miller, Thomas F},
  journal={The Journal of chemical physics},
  volume={153},
  number={12},
  year={2020},
  publisher={AIP Publishing}
}

@article{unke2019physnet,
  title={PhysNet: A neural network for predicting energies, forces, dipole moments, and partial charges},
  author={Unke, Oliver T and Meuwly, Markus},
  journal={Journal of chemical theory and computation},
  volume={15},
  number={6},
  pages={3678--3693},
  year={2019},
  publisher={ACS Publications}
}

@article{schutt2017quantum,
  title={Quantum-chemical insights from deep tensor neural networks},
  author={Sch{\"u}tt, Kristof T and Arbabzadah, Farhad and Chmiela, Stefan and M{\"u}ller, Klaus R and Tkatchenko, Alexandre},
  journal={Nature communications},
  volume={8},
  number={1},
  pages={13890},
  year={2017},
  publisher={Nature Publishing Group UK London}
}

@article{smith2017ani,
  title={ANI-1: an extensible neural network potential with DFT accuracy at force field computational cost},
  author={Smith, Justin S and Isayev, Olexandr and Roitberg, Adrian E},
  journal={Chemical science},
  volume={8},
  number={4},
  pages={3192--3203},
  year={2017},
  publisher={Royal Society of Chemistry}
}

@article{frank2022so3krates,
  title={So3krates: Equivariant attention for interactions on arbitrary length-scales in molecular systems},
  author={Frank, J Thorben and Unke, Oliver T and M{\"u}ller, Klaus-Robert},
  journal={arXiv preprint arXiv:2205.14276},
  year={2022}
}

@article{unke2021spookynet,
  title={SpookyNet: Learning force fields with electronic degrees of freedom and nonlocal effects},
  author={Unke, Oliver T and Chmiela, Stefan and Gastegger, Michael and Sch{\"u}tt, Kristof T and Sauceda, Huziel E and M{\"u}ller, Klaus-Robert},
  journal={Nature communications},
  volume={12},
  number={1},
  pages={7273},
  year={2021},
  publisher={Nature Publishing Group UK London}
}

@article{ko2021fourth,
  title={A fourth-generation high-dimensional neural network potential with accurate electrostatics including non-local charge transfer},
  author={Ko, Tsz Wai and Finkler, Jonas A and Goedecker, Stefan and Behler, J{\"o}rg},
  journal={Nature communications},
  volume={12},
  number={1},
  pages={398},
  year={2021},
  publisher={Nature Publishing Group UK London}
}

@article{gasteiger2021gemnet,
  title={Gemnet: Universal directional graph neural networks for molecules},
  author={Gasteiger, Johannes and Becker, Florian and G{\"u}nnemann, Stephan},
  journal={Advances in Neural Information Processing Systems},
  volume={34},
  pages={6790--6802},
  year={2021}
}

@article{bartok2010gaussian,
  title={Gaussian approximation potentials: The accuracy of quantum mechanics, without the electrons},
  author={Bart{\'o}k, Albert P and Payne, Mike C and Kondor, Risi and Cs{\'a}nyi, G{\'a}bor},
  journal={Physical review letters},
  volume={104},
  number={13},
  pages={136403},
  year={2010},
  publisher={APS}
}

@article{deringer2021gaussian,
  title={Gaussian process regression for materials and molecules},
  author={Deringer, Volker L and Bart{\'o}k, Albert P and Bernstein, Noam and Wilkins, David M and Ceriotti, Michele and Cs{\'a}nyi, G{\'a}bor},
  journal={Chemical reviews},
  volume={121},
  number={16},
  pages={10073--10141},
  year={2021},
  publisher={ACS Publications}
}

@article{christensen2020fchl,
  title={FCHL revisited: Faster and more accurate quantum machine learning},
  author={Christensen, Anders S and Bratholm, Lars A and Faber, Felix A and Anatole von Lilienfeld, O},
  journal={The Journal of chemical physics},
  volume={152},
  number={4},
  year={2020},
  publisher={AIP Publishing}
}

@article{behler2007generalized,
  title={Generalized neural-network representation of high-dimensional potential-energy surfaces},
  author={Behler, J{\"o}rg and Parrinello, Michele},
  journal={Physical review letters},
  volume={98},
  number={14},
  pages={146401},
  year={2007},
  publisher={APS}
}

@article{zhang2003quantum,
  title={Quantum Monte Carlo method using phase-free random walks with Slater determinants},
  author={Zhang, Shiwei and Krakauer, Henry},
  journal={Physical review letters},
  volume={90},
  number={13},
  pages={136401},
  year={2003},
  publisher={APS}
}


\end{document}